\documentclass[11pt]{article}
\usepackage[T1]{fontenc}
\usepackage[utf8x]{inputenc}
\usepackage{lmodern, microtype}
\usepackage[usenames,dvipsnames,svgnames,table]{xcolor}
\usepackage{graphicx,overpic}
\usepackage[framemethod=TikZ]{mdframed}
\usepackage{amsmath,amssymb,mathrsfs,mathtools,nccmath,relsize}
\usepackage{amsthm,mathrsfs,dsfont,cases,nicefrac,bm,blkarray,slashed}
\usepackage{booktabs,multirow,calc}
\usepackage{enumitem,caption}
\usepackage{clipboard,verbatim}
\usepackage{sepfootnotes}
\usepackage{cite} 
\usepackage[parsep]{collref} 
\usepackage[compat=1.1.0]{tikz-feynman}
\usepackage{ifdraft}
\ifdraft{
	\usepackage[notref,notcite]{showkeys}
	\usepackage[obeyspaces,hyphens,spaces]{url}
}{}
\usepackage{marginnote}
\usepackage[a4paper,
            hmargin =2.8cm,
            top=3cm,
            bottom=4cm,
            marginparsep=5mm,
            marginparwidth=25mm
            ]{geometry} 

\usepackage{url}
\usepackage{hyperref}
\hypersetup{linktocpage=true,colorlinks=true,linkcolor=nicecolor,citecolor=nicecolor,urlcolor=nicecolor,final}
\hypersetup{bookmarksnumbered=true,bookmarksopenlevel=1}
\definecolor{nicecolor}{rgb}{0.1, 0.3, 0.4}

\newcommand{\const}{c}

\newcommand{\mn}{m_n}
\newcommand{\mkk}{m_{\rm \textsc{kk}}}
\newcommand{\mcb}{m_{\rm \textsc{cb}}}

\renewcommand{\a}{\alpha}
\renewcommand{\b}{\beta}
\newcommand{\g}{\gamma}

\renewcommand{\o}{\omega}

\newcommand{\e}{\epsilon}

\newcommand{\s}{\sigma}

\newcommand{\f}{{\mathsf{f}}}

\newcommand{\cubN}{\c F}
\newcommand{\inner}{\hspace{-1pt}
	\raisebox{1.1pt}{\scalebox{0.8}{
			$\mathsmaller{\mathsmaller{\bullet}}\,\,$}}\hspace{-1pt}}

\newcommand{\innerY}{\raisebox{1.5pt}{$\,\mathsmaller{\mathsmaller{\circ}}\,$}}

   \renewcommand{\S}{\Sigma}

\renewcommand{\c}{\mathcal}

\newcommand{\cV}{\mathcal{V}}

\newcommand{\reg}{{\rm reg}}
\newcommand{\regb}{{\rm \textbf{reg}}}

\newcommand{\we}{\wedge}
\newcommand{\wt}[1]{\widetilde{#1}}
\newcommand{\ws}{\wedge \star}
\newcommand{\wsh}{\wedge \hat \star}

\DeclareFontFamily{U}{mathx}{\hyphenchar\font45}
\DeclareFontShape{U}{mathx}{m}{n}{<-> mathx10}{}
\DeclareSymbolFont{mathx}{U}{mathx}{m}{n}
\DeclareMathAccent{\widebar}{0}{mathx}{"73}

\newcommand{\hatj}{\hat{\jmath}\,}

\newcommand{\rb}{{\rm b}}

\newcommand{\jh}{{\hat \jmath}}

\newcommand{\oneloop}{\text{1-loop}}
\newcommand{\cs}{{\rm \textsc{cs}}}
\newcommand{\gs}{{\rm \textsc{gs}}}

\newcommand{\vzt}[1]{\vev{\zeta^{#1}}}
\newcommand{\Z}{\mathbb Z}

\newcommand{\intdkn}[1]{\int \frac{\d^{#1}k}{(2\pi)^{#1}}}

\newcommand{\intl}{\int \frac{\d^5l}{i\pi^3 }}

\newcommand{\limm}{\lim_{M\to \infty}}

\newcommand{\sumn}{\sum_{n \in \Z}}

\newcommand{\mpl}[1]{M_{\rm pl,#1}}

\newcommand{\hoo}[1]{h^{1,1}(#1)}

\newcommand{\K}[1]{\c K_{#1}}

\newcommand{\nvf}{n_V^{(5)}}
\newcommand{\nv}{n_V}
\newcommand{\nh}{n_H}
\newcommand{\nt}{n_T}

\newcommand{\SO}{SO(1,\nt)}
\newcommand{\class}{{\rm class}}

\newcommand{\susy}[1]{$\c N = #1$}

\definecolor{myred}{rgb}{0.7, 0.11, 0.11}
\colorlet{tableheadcolor}{gray!15} 
\colorlet{tablerowcolor}{gray!7} 
\newcommand{\note}[1]{\ifdraft{\textcolor{DarkRed}{\hspace{1mm}[#1] }\marginnote[$\Rightarrow$]{$\Leftarrow$}{\reversemarginpar \marginnote[$\Rightarrow$]{$\Leftarrow$}}}{}}

\newcommand{\mytodo}[1]{}
\renewcommand{\comment}[1]{\iffalse #1	\fi}

\newcommand{\msplit}[1]{\hspace{#1 mm} \\}

\DeclareMathOperator{\dd}{d \!}
\DeclareMathOperator{\sign}{sign}

\DeclareMathOperator{\tr}{tr}
\renewcommand{\d}{\dd \, \!}

\newcommand{\vev}[1]{\langle #1 \rangle}
\newcommand{\floor}[1]{\lfloor #1 \rfloor}

\newcommand{\mmfrac}[2]{\mfrac{\raisebox{-1pt}{$#1$}}{#2}}

\newcounter{listcount}\newcounter{totalcount}%
\newcommand{\ideal}[1]{%
	\setcounter{totalcount}{0}
	\renewcommand*{\do}[1]{\stepcounter{totalcount}}
	\docsvlist{#1}
	\setcounter{listcount}{0}
	\renewcommand*{\do}[1]{
		\stepcounter{listcount}
		##1\ifnum\value{listcount}<\value{totalcount}\,;\fi
	}
	\left(\docsvlist{#1}\right)
}

\newcommand{\qqtext}[1]{\qquad \text{#1} \qquad}

\numberwithin{equation}{section}
\begin{document}

\newpage

\vspace*{-2.5cm}
\begin{flushright}    
  {\small 
   IFT-UAM/CSIC-20-076
  }
\end{flushright}

\vspace{2cm}
\begin{center}        
  {\LARGE 6d $\c N =  (1,0)$ anomalies on $ S^1 $ \\[5mm] and F-theory implications}
\end{center}

\vspace{1cm}
\begin{center}        
{\large  Pierre Corvilain}
\end{center}

\vspace{0.1cm}
\begin{center}        
\emph{Instituto de F\'isica Te\'orica UAM-CSIC, \\[2mm] Cantoblanco, 28049 Madrid, Spain}\\[0.15cm]
\end{center}

\vspace{0.1cm}
\begin{center}
\texttt{pierre.corvilain @ uam.es}
\end{center}

\vspace{1.5cm}

\begin{abstract}

\noindent
We show that the pure gauge anomalies of 6d \susy{(1,0)} theories compactified on a circle are captured by field-dependent Chern-Simons terms appearing at one-loop in the 5d effective theories. These terms vanish if and only if anomalies are canceled. In order to obtain this result, it is crucial to integrate out the massive Kaluza-Klein modes in a way that preserves 6d Lorentz invariance; the often-used zeta-function regularization is not sufficient. 
Since such field-dependent Chern-Simons terms do not arise in the reduction of M-theory on a threefold, six-dimensional F-theory compactifications are automatically anomaly free, whenever the M/F-duality can be used.
A perfect match is then found between the 5d $\c N = 1$ prepotentials of the classical M-theory reduction and one-loop circle compactification of an anomaly free theory.
Finally, from this potential, we read off the quantum corrections to the gauge coupling functions.

\end{abstract}
\thispagestyle{empty}
\clearpage

\setcounter{tocdepth}{2}

\tableofcontents

\vspace*{0.5cm}

\section{Introduction}

For a theory to be consistent, it cannot have anomalies in its local conservation laws, that is, gauge invariance or general covariance. 
This can lead to very stringent constraints on gravitational theories, especially in $4k+2$ dimensions, as Alvarez-Gaumé and Witten showed~\cite{AlvarezGaume:1983ig} that in those dimensions one can have pure gravitational anomalies, on top of the usual gauge~\cite{Adler:1969gk, Bell:1969ts} 
and mixed~\cite{Delbourgo:1972xb} ones.
For instance this selects unique ten and six-dimensional supergravities with \susy{(2,0)} and only two with \susy{(1,0)} in 10d.
Six-dimensional \susy{(1,0)} supergravities are more interesting as, while their anomaly cancellation conditions are still very restrictive, they allow for a wide range of gauge groups and matter representations~\cite{Schwarz:1995zw,Kumar:2009ae,Kumar:2009us,Kumar:2010ru,Kumar:2010am,Seiberg:2011dr,Park:2011wv,Taylor:2019ots}. 
In most cases, and in particular whenever there are hypermultiplets charged under $ U(1) $ gauge fields, the cancellation of anomalies requires a 6d analog~\cite{Green:1984bx,Sagnotti:1992qw} of the 10d Green-Schwarz mechanism~\cite{Green:1984sg}, which consists in adding a non-invariant classical in order to cancel the one-loop variation.
In this note we ask ourselves the question: what happens to 
such abelian 6d \susy{(1,0)} theories when compactified on a circle?
In particular, how is the Green-Schwarz mechanism translated in one dimension lower: there will still be a non-invariant classical piece, but is it still canceled by a non-invariant one-loop piece? 
Another way of phrasing this is to ask what happens when compactifying on a circle a 6d theory whose anomalies are not canceled: since odd-dimensional theories are non-chiral and thus do not suffer from local anomalies, are the 6d anomalies disappearing upon compactification?

The answer is naturally \emph{no}, anomalies are not lost upon compactification and, when the Green-Schwarz mechanism is at work, one-loop variations are still canceled by non-invariant classical pieces.
If anywhere, one should look at the CS terms, which violate parity, in order to find out where the anomalies are captured in 5d. 
And indeed, after integrating out (in a proper way, as we will see below) the KK-towers of the fields leading to the 6d anomalies, we find that non-gauge invariant CS terms appear at one-loop, as their `coefficients' then depend on other fields. The variation of these terms, both under gauge transformations and shifts of the scalars, precisely reproduce the expected variations from the 6d anomalies. When the latter are canceled by a Green-Schwarz mechanism, the circle reduction of the classically non-invariant piece also produce such field-dependent CS terms, which exactly cancel with the one-loop ones. We thus see that the 5d theory is consistent if and only if the 6d one also is; in other words, the anomaly is preserved under circle compactification.

A crucial ingredient for obtaining this result is the regularization one uses in order to integrate out the massive modes: we find that it needs preserve \emph{6d Lorentz invariance}. To understand why, note that the five dimensional theory that we are considering, is not \emph{any} five-dimensional theory, rather it comes from a circle reduction, meaning that in the UV it becomes six-dimensional.\footnote{At length scales much smaller than the inverse radius, one cannot distinguish between $\mathbb R^{1,4} \times S^1$ and~$\mathbb R^{1,5}$.}
This constrains the five-dimensional theory, at least when it comes to anomalies, as they are due to the lack of regulator preserving \emph{simultaneously} gauge invariance, or general covariance, and Lorentz invariance \emph{in the UV}.
Requiring the 5d regulator to also respect 6d Lorentz invariance, one is equally forced to give up gauge invariance, or general covariance, and  anomalies are preserved. 
Another way of seeing this is to recall that odd-dimensional theories do not have relevant anomalies because one can always add appropriate counterterms to cancel their one-loop variation. However if the theory is secretly higher-dimensional, one is not free to add \emph{any} counterterm, they should respect the higher Lorentz invariance. 
Such considerations about the importance of the regulator in KK theories were already made in~\cite{Bauman:2007rt,Bauman:2008rr}, and exploited in~\cite{Corvilain:2017luj} to study 4d chiral anomalies on a circle. In this note we investigate the six-dimensional equivalent, where it is even more crucial, as the anomaly can only be canceled by a GS mechanism.\footnote{The GS mechanism can also be implemented in 4d, but does \emph{not have to}, as one can cancel all the anomalies by considering several fermions with appropriate $ U(1) $ charges. In that case, a regularization preserving solely three-dimensional Lorentz invariance will yield the same results (see~\cite{Corvilain:2017luj} for more details).}
In some sense the argument can be understood as `local anomalies are also preserved along the RG flow', thus a similar argument should hold for any dimension or compactification space.

Concretely, we will focus on pure gauge anomalies of six-dimensional \susy{(1,0)} supergravities. For clarity we will take the gauge group to be abelian, but the generalization to non-abelian gauge groups should be straightforward, as they are anyway broken down to their Cartan subgroup in the Coulomb branch. 
We will keep the number of $ U(1) $'s arbitrary\footnote{Although the number $N$ of abelian gauge factors can in principle be infinite, as far as anomalies are concerned~\cite{Taylor:2011wt}, one can show~\cite{Lee:2019skh}, using the completeness hypothesis and anomaly inflow techniques, that for the theory not to belong to the Swampland, one should have $N \leq 20$.} and denote them by $A^i$, while $A^0$ will denote the graviphoton.
We then find the pure gauge anomaly to arise at one-loop (using an adequate regularization, as mentioned earlier) in the 5d CS coefficients $k_{ijk}$, $k_{0ij}$ and $k_{00i}$, as they all have a field-dependent pieces which cancel if only if the GS mechanism is at work (see section~\ref{sec:6dS1} for details). The fact that the pure gauge anomaly is encoded in $k_{ijk}$ was already suspected in~\cite{Bonetti:2011mw,Grimm:2013oga} and made more precise in~\cite{Grimm:2015zea} using shift-symmetries of the Coulomb branch parameters, which originate from the large gauge transformations along the circle. In this note, we fully clarify how the anomaly comes about in all the CS terms involving an $A^{i}$, and not only in $k_{ijk}$. Importantly, this allows us to find the full 5d \susy{1} (8 supercharges) prepotential for this circle reduction, and we find that it is a cubic function if and only if the anomaly of the 6d theory cancel. From this prepotential, we infer the quantum corrections to the kinetics terms.

\sepfootnotecontent{b}{The F-theory set-up is in principle recovered when one takes the fiber volume to vanish. This sends both the radius of the F-theory circle and the Coulomb branch parameter to zero, making infinitely many fields to become light. While this indicates the extra dimension, it is hard to implement in practice. Rather the match is done at finite radius and in the Coulomb branch.}

In the second part of this note we investigate the consequences of our anomaly analysis for 6d \susy{(1,0)} theories obtained as compactifications of F-theory on elliptically fibered Calabi-Yau threefolds~\cite{Vafa:1996xn,Morrison:1996na,Morrison:1996pp}.
In order to determine the effective action, we use the M/F-duality~\cite{Denef:2008wq}, which tells us that M-theory reduction is dual to the circle compactification of the F-theory one.
More precisely, the \emph{classical} reduction of  eleven-dimensional supergravity on the resolved Calabi-Yau threefold (if a resolution exists, that is) is to be compared, in the correct duality frame, with the circle reduction of a generic six-dimensional \susy{(1,0)} theory, \emph{including quantum corrections}~\cite{Witten:1996md,Intriligator:1997pq}.
More specifically, it is essential to push the latter to the Coulomb branch and integrate out all massive modes, including all KK-modes, which was noticed in~\cite{Bonetti:2011mw} and subsequently implemented in~\cite{Bonetti:2012fn, Bonetti:2013ela}.\sepfootnote{b} One might be worried that information contained in the massive modes is lost. While this is in general true, some information is nonetheless captured by the quantum corrections to the zero modes. In particular, as explained above, all the knowledge about anomalies is maintained, as they are captured in 5d by field-dependent CS terms, at least when it comes to the pure gauge anomaly.
Such terms are absent from the M-theory reduction, which can therefore only be matched with the circle compactification of theory without pure gauge anomalies.
This implies that all 6d $ \c N = (1,0) $ F-theories obtained via the M/F-duality are automatically free of gauge anomalies. This strategy was already used in~\cite{Corvilain:2017luj} to show a similar result for 4d \susy{1} F-theories.
Subsequently, one would like to go further and match every term of both sides of the duality, i.e.~the M-theory reduction and the circle compactification of an anomaly free theory. 
This match was initiated in~\cite{Ferrara:1996wv,Bonetti:2011mw} and continued in~\cite{Grimm:2013oga} using the one-loop results of~\cite{Bonetti:2013ela}, however a precise match of all the terms was never achieved due the non-invariant Green-Schwarz CS terms which are left unmatched if one does not compute the one-loop corrections using an adequate  regularization, in the sense explained earlier. With our method, we are able to perfectly match every terms of both actions, on condition that also the pure gravitational and mixed gauge-gravitational anomalies are canceled.

Finally let us note that part of the elegance of F-theory is that many quantities get geometrized, and in particular also the anomaly coefficients 
\cite{Sadov:1996zm,Kumar:2009ac,Grassi:2011hq,Park:2011ji,Lee:2018ihr}.
The anomaly cancellation conditions can then in turn be translated into geometrical identities~\cite{Grassi:2011hq,Park:2011ji,Bies:2017abs}, which have been checked for specific geometries but not proved in general. Our analysis can be seen as a `physics proof' of these relations.

The rest of this note is organized as follows: in section~\ref{sec:6dS1} we reduce a generic 6d \susy{(1,0)} supergravity with abelian gauge group on a circle and compute the 5d one-loop CS terms that result from integrating out the massive modes. We explain in detail how the choice of regularization is crucial for the resulting theory to be gauge invariant, but we leave the computational details for appendix~\ref{app:loop}. 
In section~\ref{sec:Mthy} we present the classical reduction of M-theory on an elliptically Calabi-Yau threefold, and compare it with the circle compactification, finding a perfect match.

\section{\boldmath 6d $\c N = (1,0)$ supergravity on $S^1$}
\label{sec:6dS1}

In this section we perform the circle reduction of a generic six-dimensional supergravity with (1,0) supersymmetry (8 supercharges), restricting to abelian gauge group for simplicity. We first explain the field content and how anomalies are canceled via the GS mechanism. Then we perform the classical circle reduction, mostly following the notation of~\cite{Bonetti:2011mw,Grimm:2013oga}. Finally, we compute the one-loop effective action by integrating all the massive KK-modes, paying special attention to different types of regularizations.

\subsection{Field content and anomaly cancellation}

In a generic 6d \susy{(1,0)} there can be four types of multiplets (for spins less or equal to two): the gravity multiplet, tensor multiplets, vector multiplets and hypermultiplets, whose field content is given in table~\ref{multiplets}.
\begin{table}[h!]
\begin{center}
{\def\arraystretch{1.4} \tabcolsep=15pt
\begin{tabular}{@{}llr@{}}
	\toprule
	Multiplet & Field content                                              &  Number \\
	\midrule
	Gravity   & $\hat g_{\mu\nu}, \, \hat \psi_\mu^+,\, \hat B_{\mu\nu}^+$ &       1 \\
	Tensor    & $\hat B_{\mu\nu}^-,\,\hat \chi^-,\,\hatj$                  & $ n_T $ \\
	Vector    & $\hat A_{\mu}, \,\hat \lambda^+$                           &   $n_V$ \\
	Hyper     & $\hat \psi^-,\, 4 \,\hat  q$                               &   $n_H$ \\
	\bottomrule
\end{tabular}
}
\end{center}
\caption{Field content of the different multiplets in 6d (1,0) supergravities. $\hat g_{\mu \nu}$ denotes the graviton, the $\hat B_{\mu \nu}$ are two-tensors, $\hat \psi_\mu$, $ \hat \chi $, $ \hat \lambda $ and $\hat \psi$ are Weyl spinors (respectively the gravitino, tensorini, gaugini and hyperini), $\hatj$ and $\hat q$ are scalars, and finally $ \hat A_\mu $ is a vector. The signs on the fermions indicate their chirality, while those on the two-forms $\hat B$ indicate (anti)-self-duality. We use hats on the fields to show that they are six-dimensional; later on, five-dimensional quantities will be unhatted.}
\label{multiplets}
\end{table}
The $\nt$ scalars of the tensor multiplets parametrize the coset space $\SO/SO(n_T)$, such that they are conveniently combined into an $\SO$ vector $\hatj^\a$, $\a=1,\ldots,\nt +1$ of unit norm 
\begin{equation}\label{jj}
\hatj \inner \hatj \equiv \Omega_{\a\b} \, \hatj^ \a \hatj^\b = 1 \,,
\end{equation}
with respect to a constant $\SO$ metric  $\Omega_{\a \b}$. Products made using this metric will be denoted by the dot `$\inner$' throughout this document.
It is also convenient to group the self-dual two-form $\hat B^+$ and the $\nt $ anti-self-dual two-forms $\hat B^-$ into a single $\hat B^\alpha, \ \a=1, ...,n_T+1$.

Since (1,0) theories are chiral, a generic theory is potentially anomalous. The anomaly is captured (via the descent equations) by an 8-form polynomial, which reads~\cite{Erler:1993zy,Park:2011wv}
\begin{align}\label{I8}
\begin{split}
I_8 =  - \mmfrac{1}{5760} (\nh - \nv + 29 \, \nt -273) \Big[ \tr \hat {\c R}^4  + \mmfrac{5}{4} (\tr \hat {\c R}^2)^2\Big] - \mmfrac{1}{128} (9-\nt) (\tr \hat {\c R}^2)^2 \msplit{6} + \sum_{\f=1}^{n_H^*} \, \bigg[ \, q^\f_i \,q^\f_j  \, \hat F^i \wedge \hat F^j \wedge  \left(\mmfrac{1}{96} \,  \tr \hat {\c R}^2  \, - \mmfrac{1}{24} \, q^\f_k \,q^\f_l \,  \hat F^k \wedge\hat F^l\right)\bigg]\,,
\end{split}
\end{align}
where $\hat {\cal R}$ is the curvature two-form, the $\hat F^i$'s are the field strength of the $U(1)$ gauge fields $\hat A^i$ ($i = 1, ..., n_V$), and the $q_i^\f$'s are the charges of the hyperini under the $\hat A^i$'s (with $\f=1,...,n_H^*$ denoting the number of charged hypermultiplets).
\note{Make sure all this is correct}

In order to cancel this anomaly ---i.e.~have a consistent theory--- one needs to implement a 6d version of the Green-Schwarz (GS) mechanism~\cite{Green:1984bx,Sagnotti:1992qw}, which requires the anomaly polynomial to factorize as
\begin{align}\label{fact}
I_8 = \mmfrac 12 \, \hat X_4 \inner \hat X_4 \,, \qquad \hat X_4^\a = \mmfrac 12  a^\a \tr \hat {\c R}^2 + 2 \, b_{ij}^\a \, \hat  F^i \wedge \hat F^j \,, 
\end{align}
for some $\SO$ vectors $a^\a$ and $b^\a_{ij}$, known as the anomaly coefficients. All anomalies can then be canceled by adding to the action the local counter-term 
\begin{equation} \label{GSterm}
S^\gs = -\mmfrac 12 \int_{\c M_6} \hat B \inner \hat X_4\,.
\end{equation}
Indeed, if the two-forms fields $\hat B^\a$ non-trivially transform under a Lorentz transformation $\delta \hat \omega = \d \hat l + [\hat \o, \hat l]$ and a gauge transformation $\delta \hat A^i = \d \hat \lambda^i$
\begin{equation}
\delta \hat B^\a = - \mmfrac 12 a^\a \tr \hat l \d \hat \o - 2 \, b^\a_{ij} \, \hat \lambda^i \, \hat F^j\,,
\end{equation}
the term~\eqref{GSterm} is not invariant (at classical level) and its variation precisely cancels with the one-loop one. The gauge invariant field strengths for the two-forms $\hat B^\a$ are given by
\begin{equation}\label{key}
\hat G^\a = \d \hat B^\a  + \mmfrac 12 \, a^\a \omega_{\rm grav} + 2 \, b_{ij}^\a \, \hat A^i \we \hat  F^j\,,
\end{equation}
such a form being actually required by supersymmetry.
For the factorization~\eqref{fact} to occur, and therefore the theory to be free of anomalies, the following \emph{anomaly cancellation conditions} need to hold 
\begin{subequations}\label{AC_6d}
	\begin{align}
n_H-n_V &=  273 - 29 \, \nt  \,,  \label{grav_irr} \\
a \inner a &= 9 - n_T\,, \label{grav_red} \\
a \inner b_{ij} &= - \mmfrac 16 \sum_\f  \, q^\f_i \,q^\f_j \,, \label{mixed} \\
b_{ijkl} &= \sum_\f  \, q^\f_i \,q^\f_j\, q^\f_k\, q^\f_l \,, \label{gauge}
	\end{align}
\end{subequations}
where $b_{ijkl} \equiv b_{ij} \inner b_{kl} + b_{ik} \inner b_{jl} + b_{il} \inner b_{jk}$.
The first two equations ensure the cancellation of the pure gravitational anomalies, the third one of the mixed gauge-gravity anomaly and the last one of the pure gauge anomaly.

With anomalies canceled in this way, one can write a supersymmetric action which is invariant at one-loop, whose bosonic part reads
\begin{equation}\label{6Dact}
\begin{split}
S_6 =  \mfrac 12 \, \mpl{6}^4 \int_{\c M_{6}}  R_6 \, \hat \star \,1 -  h_{UV} \d \hat q^U \wsh \, \d \hat q^V -  \hat g_{\a\b} \, \Big(\d j^\a \wsh \, \d j^\b + \mmfrac 12 \, \hat G^\a \wsh \,\hat G^\b\Big) \msplit{3}
- 2\, b_{ij} \inner j\, \, \hat F^i \wsh \,\hat F^j
-  \hat B \we \inner \Big( \, \mmfrac 12 \, a \, \tr \, \hat{\c R}^2 
+ 2 \, b_{ij} \, \hat F^i \wedge \hat F^j \Big) \,,
\end{split}
\end{equation}
where the metric for the tensor multiplets is given by $\hat g_{\a \b} = 2 \, \jh_\a \jh_\b - \Omega_{\a\b}$. This is in fact a pseudo-action as the kinetic terms of the two-forms should vanish because of their (anti)-self-duality conditions
\begin{equation}\label{duality}
\hat g_{\a\b} \, \hat \star \, \hat G^\a = \Omega_{\a \b} \, \hat G^\a\,.
\end{equation}
The easiest way out, which suffices for our purposes, is to impose these self-duality conditions at the level of the equations of motion.

\subsection{Classical circle reduction}

We now perform the circle reduction of the action~\eqref{6Dact}. Taking the circle to be of radius $r$ and along the $y$ direction, we decompose the 6d metric as
\begin{equation}\label{key}
\d \hat s^2 = \d s ^2 - r^2 (\d y - A^0)^2\,, 
\end{equation}
where $A^0$ is the graviphoton (or KK-photon).
Concerning the gauge fields, we make the following Ansätze
\begin{subequations}
	\begin{alignat}{1}\label{AnsatzB}
\hat A^i &= A^i + \zeta^i \,(\d y - A^0)\,, \\
\hat B^\a &= 4 \, B^\a - \big ( 4 \, A^\a - 2 \, b^\a_{ij} \, \zeta ^i A^j \big ) \wedge (\d y - A^0)\,,
\end{alignat}
\end{subequations}
where the $A$'s and $B$'s are 5d one-forms and two-forms respectively and the $\zeta$'s are 5d scalars.\footnote{Recall our convention that hatted fields are six-dimensional while unhatted ones are five-dimensional.}
In five dimensions, a two-form is dual to a one-form, and, due to the self-duality condition~\eqref{duality}, the $B^\a$'s are dual to the $A^\a$'s using the reduction of~\eqref{duality} on the circle. After this dualization, the tensor multiplets become vector ones in five dimensions, which is the correct duality frame to match with the M-theory reduction as we will see later.
We are now almost ready to display the resulting action in 5d, but first we need to perform a Weyl rescaling $ g_{\mu \nu}^{\rm new} = (r/r_0 )^{2/3} g_{\mu \nu} $ in order to bring the action to the Einstein frame. For later convenience, let us accompany this by a rescaling of all the vectors $  A^{\rm new} =  \const \, r_0^{-1/3} A$, with $\const =32^{1/3}$.

Altogether the five-dimensional two-derivatives action reads
\begin{equation}\label{S5class}
S_5^\class = \mpl{5}^3 \int_{\c M_{5}} \mmfrac 12\, R_5 \star 1 + \c L^{\rm kin} + \c L^{\rm gi} + \c L^{\rm ngi} \,,
\end{equation}
where the kinetic terms are given by
\begin{align}
\begin{split}
\label{Lkin}
\c L^{\rm kin} =& \ - \mmfrac 23 \, r^{-2} \, \d r \ws \d r 
- \mmfrac 14  \, r^{8/3} \wt F^0 \ws \wt F^0  -  h_{UV} \d  q^U \ws \d q^V\msplit{0}
& \hspace{-11mm} - \mmfrac 12 \, g_{\a\b} \, \Big( \d j^a \ws \d j^\b + r^{-4/3}  \wt F^\a \ws \wt  F^\b \Big) 
-2 \, j \inner b_{ij} \, \Big ( r^{-2} \d \zeta^i \ws \d \zeta^j + r^{2/3} \wt F^i \ws \, \wt F^j \Big ) ,
\end{split} \\
\intertext{the gauge invariant Chern-Simons terms by
}
\c L^{\rm g i} =&  - \mmfrac 14 A^0  \we F \we  \! \inner  F + \mmfrac 14 \, b_{ij} \inner A\we    F^i   \we F^{j} \,, \\
\intertext{and finally the non gauge invariant Chern-Simons terms by}
\c L^{\rm n g i} =&    - \mmfrac{1}{16} \, b_{ijkl}\,  \Big(  \zeta^i  A^j   \we F^k   \we  F^l -  \, \zeta^i \zeta^j A^k \we   F^l  \we F^0 + \mfrac 13 \,  \zeta^i \zeta^j \zeta^k A^l  \we  F^0  \we   F^0 \Big) \,. \label{ngi}
\end{align}
In~\eqref{Lkin} the field strengths are given by 
\begin{equation}\label{curlyF}
\begin{alignedat}{2}
&\wt F^0 & &=  c^{-1} \, F^0 \,, \\
&\wt F^i & &=    c^{-1} \, \big (F^i - \zeta^i  F^0 \big ) \,,\\
&\wt F^\a & &=  c^{-1}\, \Big[4 \,F^\a +2 \, b_{ij}^\a \,  \zeta^i \, \big  (\zeta^j F^0 - 2 F^j \big )\Big] \,.
\end{alignedat}
\end{equation}
This classical action is non gauge invariant, as can directly be seen by the field-dependent CS terms in~\eqref{ngi}. This is perfectly normal, as it is the classical reduction of~\eqref{6Dact}, which is also non gauge invariant at tree-level, due to the Green-Schwarz terms. Only at one-loop is the 6d theory gauge invariant, and the same holds in 5d as well, as we will shortly clarify.

In particular, this means that one should not expect to be able to write~\eqref{S5class} in the canonical supersymmetric form, that is, as 
\begin{equation}\label{S5can}
\begin{split}
S_5^{\rm can} = \mfrac 12 \, \mpl{5}^3 \int_{\c M_{5}} R_5 \star 1 -  h_{uv} \d  q^u \ws \d q^v 
\msplit{60}
- G_{IJ} \, \Big(\d \phi^I \ws \d \phi^J + \bar F^I \ws \, \bar  F^J\Big)  
- \mmfrac{1}{6} \, k_{IJK} \bar  A^I \wedge \bar F^J \wedge \bar F^K \, ,
\end{split}
\end{equation}
for this is manifestly gauge invariant. In~\eqref{S5can}, the kinetic and CS terms of the vector multiplets are both determined by a single cubic function, known as the prepotential, defined as
\begin{equation}\label{cubN}
 \cubN = \mfrac{1}{3!} \, k_{IJK} \, \phi^I \phi^J \phi^K\,, 
\end{equation}
subject to the constraint 
\begin{equation}\label{constr}
\cubN =1\,.
\end{equation}
This defines a \emph{very special geometry}, whose metric is given by 
\begin{equation}\label{gcf}
G_{IJ} =  \left .- \frac 12 \frac{ \partial^2 \log \cubN}{\partial \phi^I \partial \phi^J} \right|_{\cubN =1} \,.
\end{equation}
Regarding the CS terms, they are evidently obtained as $ k_{IJK} = \partial_I \partial_J \partial_K \cubN \equiv (\cubN)_{IJK}$.

Nevertheless, it turns that out there is a way to cast the action~\eqref{S5class} in a form similar to that of~\eqref{S5can}. 
To start with, one splits the canonical 5d coordinates as $\phi^I = (\phi^0,\phi^\a,\phi^i)$ and identifies them with the ones coming from the circle reduction as\footnote{This seems to be an arbitrary choice for now, as there are many other choices that do the job. However this particular choice is singled out by the comparison with the M-theory reduction, as we will see in section~\ref{sec:Mthy}.} 
\begin{subequations} \label{very_spec_coord}
	\begin{alignat}{3}	
	    &\phi^0 &&= \,\const  \,            && r^{-4/3}  \,,\\
		&\phi^\a  &&=  \frac{\const}{4}\, && r^{2/3} \,\Big (j^\a + 2 \, r^{-2} \, b^\a_{ij} \, \zeta^i \zeta^j \Big ) \,, \\
		&\phi^i &&= \, \const \,             && r^{-4/3} \, \zeta^i \,.
	\end{alignat}
\end{subequations}
Then one considers the following prepotential
\begin{align} \label{Nclass}
\cubN_{S^1}^\class = \cubN^{\rm p} + \cubN^{\rm np} \qqtext{where} 
\arraycolsep=1pt
\left \{ \begin{array}{ll}
\cubN^{\rm p} &=  \mmfrac{1}{2} \, \phi^0\,  \phi \inner \phi - \mmfrac{1}{2}   \, b_{ij} \inner  \phi \, \phi^i \phi^j \,, \\[2.5mm]
\cubN^{\rm np} &= \mmfrac{1}{24} \, \displaystyle  b_{ijkl}   \, \frac{\phi^i \phi^j \phi^k \phi^l}{\phi^0} \,,
\end{array} \right .
\end{align}
which is not cubic, although still a homogeneous polynomial of degree three. 
Using the identifications~\eqref{very_spec_coord}, one verifies that $\cubN_{S^1}^\class = j \inner j =1$ i.e.~the constraint~\eqref{constr} is satisfied, and that the kinetic terms~\eqref{Lkin} are recovered using~\eqref{gcf}. However, the Chern-Simons terms are to be computed slightly differently, namely as 
\begin{equation}\label{S5CS}
S_5^\cs = \int_{\c M_{5}} -\mmfrac 1{12}  (\cubN^{\rm p})_{IJK} \bar  A^I \we \bar  F^J \we \bar F^K -\mmfrac 1{16} (\cubN^{\rm np})_{iJK} \bar A^i \we \bar F^J \we \bar F^K \,.
\end{equation}
This is necessary in order to account for the non-gauge invariant piece~\eqref{ngi}.
As mentioned above, this non-invariance should be canceled by the one-loop corrections, which we compute next.

\subsection{One-loop corrections}

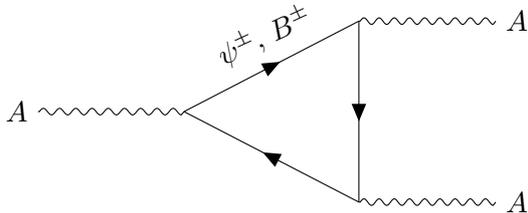
\begin{figure}[h]
\centering
\hspace{-3mm}
\begin{tikzpicture}
\begin{feynman}
\vertex (aa) {\(A\)} ; 
\vertex [right=2.2 cm of aa] (a); 
\vertex [right=2.3cm of a] (ai); 
\vertex [above= 1.2cm of ai] (b);
\vertex [below=1.2 cm of ai] (c);
\vertex [right=1.8 cm of b] (ab) {\(A\)};
\vertex [right=1.8 cm of c] (ac) {\(A\)};
\vertex [right =1.2 cm of a] (b3);
\vertex [above = 0.7cm of b3, rotate=25] (a2) {\(\psi^\pm, \, B^\pm\)};

\diagram* {
(aa) -- [photon] (a),
(a) -- [fermion] (b) -- [fermion] (c) -- [fermion] (a), 
(b) -- [photon] (ab),
(c) -- [photon] (ac),
};
\end{feynman}
\end{tikzpicture}
\captionsetup{width=.59\linewidth}
\caption{Triangle diagram with three external gauge fields and fields with parity violating actions running in the loop, generating five-dimensional CS terms.}
\label{fig:aaa5d-no-text}
\end{figure}

Five-dimensional Chern-Simons terms are generated at one-loop\footnote{And by usual non-normalization arguments, they do not receive further corrections.} by the triangle diagrams depicted on figure~\ref{fig:aaa5d-no-text}, with massive spin-1/2, spin-3/2 or tensors fields charged under the external gauge fields running in the loop
The contribution of a spin-1/2 Weyl fermion of positive chirality with mass~$m$ and charge $q$ running in the loop is~\cite{Witten:1996qb}
\begin{equation}\label{sign}
\c L^\oneloop_\cs =  \mmfrac{1}{24}  \, q^3  \, \sign (m) \, A \wedge F \wedge F\,,
\end{equation}
in our normalization. The contributions from the spin-3/2 gravitino and the self-dual tensors have been computed in~\cite{Bonetti:2013ela}; they are similar to~\eqref{sign}, with an overall factor f $5$ for the gravitino and of $-2$ for the self-dual tensors. 

On the one hand, if each external gauge field is a graviphoton $A^0$, all the KK-modes of the spin-1/2, spin-3/2 and tensor fields contribute, as they are all charged under $A^0$, with charge $n$ for the $n$-th KK-mode. In addition their masses read
\begin{equation}\label{key}
\mkk = \mfrac n r\,\left(\mfrac {r_0}{r}\right)^{1/3}\,.
\end{equation}
On the other hand, if at least one of the external gauge fields is one of the $A^i$'s, solely the charged hyperini can run in the loop, as only those are charged under the $A^i$'s.
These fields acquire in the Coulomb branch (CB) a mass proportional to $q^\f_i \, \vev{\zeta^i}$, where the the $\zeta^i$'s are the scalars parametrizing the CB\footnote{They are the scalars belonging the vector multiplets, which arose in the circle reduction from the 6d vectors, see eq.~\eqref{AnsatzB}.}. The total mass of the $n$-th KK-mode of a charged hyperino $\f$ in the Coulomb branch thus reads
\begin{equation}\label{mn}
\mn^\f = \mfrac 1 r\,\left(\mfrac {r_0}{r}\right)^{1/3} \Big( n + q_i^\f \, \vev{\zeta^i}\Big)\,.
\end{equation}

Using the above information, one finds the one-loop CS terms\footnote{Here we omit the wedges for compactness of notation, a convention that 
we will often adopt in the rest of this note.}
\begin{align} \label{CSone-loop}
\c L^\oneloop_\cs & = - \mmfrac{1}{12} \, \Big (k_{ijk} A^i F^j F^k + 3\, k_{0ij} A^i F^j F^0 + 3\, k_{00i} A^i F^0 F^0 + k_{000}A^0 F^0 F^0 \Big ) \,,
\end{align}
with the following coefficients 
\begin{subequations}\label{sums}
	\begin{alignat}{2}
	k_{ijk} & =   \mmfrac 12 \sum_{n \in \Z} \sum_\f q^\f_i \,q^\f_j\, q^\f_k \sign (\mn^\f) \,,  \\
	k_{0ij} & =  \mmfrac 12\sum_{n \in \Z}  n\sum_\f q^\f_i \,q^\f_j \sign (\mn^\f) \,, \\
	k_{00i} & =  \mmfrac 12  \sum_{n \in \Z} n^2 \sum_\f q^\f_i  \sign (\mn^\f) \,, \\
	k_{000} & =   \mmfrac 12 \sum_{n \in \Z} n^3 \bigg[ \sum_{\f}  \sign (\mn^\f) + \Big( n_H^0 - n_V + 2 \, (1-n_T) - 5 \Big) \sign (\mkk) \bigg]	 \,,
	\end{alignat}
\end{subequations}
where the sums over $n$ take all the KK-modes into account. These sums are ill-defined and need to be regularized. Using the zeta function regularization, as is often done in the literature, one obtains finite CS coefficients, namely
\begin{subequations}\label{zeta}
	\begin{alignat}{2}
	k^\reg_{ijk} & =  \sum_\f q^\f_i \,q^\f_j\, q^\f_k \left( \mmfrac 12 + S_1(\mu^\f)- \, q^\f_l \, \vev{\zeta^l }\right) \,,\label{zetaijk}\\
	k^\reg_{0ij} & =  \sum_\f q^\f_i \,q^\f_j \left( - \mmfrac{1}{12}  + S_2(\mu^\f)+ \mmfrac 12 \, q^\f_k \,q^\f_l  \, \vev{\zeta^k }\vev{\zeta^l} \right) \,, \\
	k^\reg_{00i} & =   \sum_\f q_i^\f \left(S_3(\mu^\f) - \mmfrac 13  \, q^\f_j \,q^\f_k\, q^\f_l \, \vev{\zeta^j} \vev{\zeta^k} \vev{ \zeta^l}\right) \,, \\
	k^\reg_{000} & =  \sum_{\f } \Big ( S_4(\mu^\f) + \mmfrac 14 \, q^\f_i   \, q^\f_j \,q^\f_k\, q^\f_l \, \vev{\zeta^i} \vev{\zeta^j} \vev{\zeta^k} \vev{\zeta^l} \Big ) +  \mmfrac{1}{120} (n_H-n_V-n_T-3) \,,
	\end{alignat}
\end{subequations}
where we defined the functions
\begin{equation} \label{floors}
	\begin{alignedat}{2}
    S_1(x) & = \floor x \,,\qquad  & 
 S_3(x) & = \mmfrac 16 \floor x \left( \floor x + 1 \right)\left(2\floor x +1 \right),\\
 S_2(x) & =  \mmfrac{1}{2} \floor x (\floor x +1) \,,   \qquad  &  S_4(x) & =  - \mmfrac 14 \floor{x}^2 \left(\floor{x}+1\right)^2 \,,
\end{alignedat}
\end{equation}
with $\floor{\cdot}$ being the floor function\footnote{It returns the biggest integer less than or equal to its entry.}. Even though the functions $ S_{k} $ are integer-valued,\footnote{This can be seen from the facts that $ \floor x (\floor x +1) =0 \mod 2 \, $ and $ \floor x \left( \floor x + 1 \right)\allowbreak \left(2\floor x +1 \right) =0 \mod 6 $.} the regularized CS coefficients~\eqref{zeta} are not properly quantized, since they depend continuously on the Coulomb branch parameters~$\vev{\zeta^i}$. This means that the theory is not invariant under 5d large gauge transformations and is a remnant of the non gauge invariance of the one-loop part of the 6d theory it originates from.

Nonetheless, the result~\eqref{zeta} is not satisfactory, as
it does not produce the correct variation for the full action to be gauge invariant at one-loop. Indeed~\eqref{CSone-loop} with these coefficients is actually invariant under both small gauge transformations $\delta A^i = \d \lambda^i$ and shifts of the scalars $\delta \zeta^i= n^i$, which originate from 6d small and large gauge transformations. Crucially, the 6d 1PI effective action is not invariant under them, the variation being canceled by the GS mechanism. In other words, the one-loop piece~\eqref{CSone-loop} with the coefficients~\eqref{zeta} does not cancel the field-dependent CS terms~\eqref{ngi} coming from the GS mechanism.

This situation is completely analogous to the 3d one studied in reference~\cite{Corvilain:2017luj}.
As explained in the introduction, the key point is that the regularization should respect 6d Lorentz invariance, because the UV of the theory is six-dimensional, and the 
divergences precisely come from modes running in the loop with very high momenta. It turns out that the zeta function regularization does not have this property. 
In contrast, one can ensure that the regularization respects 6d Lorentz invariance by starting with a 6d PV regulator and reducing it on a circle. This yields a 5d regulator which by construction breaks gauge invariance but respects 6d Lorentz invariance. 
The details are left for appendix~\ref{app:loop}, and the results read\footnote{The bold superscript $\regb$ indicate that we used a different regularization than before.}
\begin{subequations}\label{PV}
	\begin{alignat}{2}
	k^\regb_{ijk} & =  \sum_\f q^\f_i \,q^\f_j\, q^\f_k \left( \mmfrac 12 + S_1(\mu^\f)- \mmfrac 34 \, q^\f_l \, {\zeta^l }\right) \,, \label{PVijk}\\
	k^\regb_{0ij} & =  \sum_\f q^\f_i \,q^\f_j \left( - \mmfrac{1}{12}  + S_2(\mu^\f)+ \mmfrac 14 \, q^\f_k \,q^\f_l  \, {\zeta^k }{\zeta^l} \right) \,, \\
	k^\regb_{00i} & =   \sum_\f q_i^\f \left(S_3(\mu^\f) - \mmfrac 1{12}   \, q^\f_j \,q^\f_k\, q^\f_l \, {\zeta^j} {\zeta^k} { \zeta^l}\right) \,, \\
	k^\regb_{000} & =   \sum_{\f } S_4(\mu^\f)  + \mmfrac{1}{120} (n_H-n_V-n_T-3)\,.
	\end{alignat}
\end{subequations}
Notice that these `coefficients' now depend on the full scalar fields $\zeta^i = \vzt{i} + \chi^i$, as diagrams with the scalar fluctuations $\chi^i$ as external legs do no longer vanish using this regularization. In addition the factors in front of the vevs $ \vev{\zeta^i} $ differ from those in~\eqref{zeta}.
Both these features are crucial for the field-dependent part of these one-loop terms to combine as it should with the GS terms~\eqref{ngi}, namely as
\begin{equation}\label{ngi_all}
\c L^{\rm n g i, \, quant} =   - \mmfrac{1}{16} \, \Big (b_{ijkl} - \sum_\f  \, q^\f_i \,q^\f_j\, q^\f_k\, q^\f_l \Big )\,  \Big(  \zeta^i  A^j    F^k     F^l -  \, \zeta^i \zeta^j A^k    F^l   F^0 + \mfrac 13 \,  \zeta^i \zeta^j \zeta^k A^l    F^0     F^0 \Big) \,.
\end{equation}
These non-invariant terms vanish if and only if the pure gauge anomaly~\eqref{gauge} cancels, 
making precise the earlier statement `the anomaly is conserved upon circle reduction'.

It is comforting to note that our regularization depends on the dimensionality: a similar analysis was carried out for the three-dimensional Chern-Simons term in~\cite{Corvilain:2017luj} and the factors there 
differ from those in~\eqref{PV}, even though the same sums as in~\eqref{sums} appear. This is a consequence of the fact that the regularizations used respect different Lorentz symmetries (4d and 6d respectively). On the contrary, the zeta function regularization does not share this property and yields the same factors in both cases.

In fact,~\eqref{zeta} gives the covariant effective action, while~\eqref{PV} gives the consistent one~\cite{Jensen:2013kka,DiPietro:2014bca}, and the two are related via anomaly inflow, as follows. The 6d pure gauge anomaly $\delta S_6^{\textsc{1pi}} = \frac{1}{48} \, \sum_\f  \, q^\f_i \,q^\f_j\, q^\f_k\, q^\f_l \int_{\c M_6}  \hat \lambda^i \hat F^j \hat F^k \hat F^l $ can be canceled by taking the space-time $\c M_6$ to be the boundary of some seven dimensional space $\c M_7$, and considering in the bulk a CS term
\begin{equation}\label{CS7}
S^{\rm inflow}_7 = \mfrac 1{48} \, \sum_\f  \, q^\f_i \,q^\f_j\, q^\f_k\, q^\f_l \int_{\c M_{7}} A^i_{(7)} F^j_{(7)} F^k_{(7)} F^l_{(7)} \,,
\end{equation}
where $A_{(7)}^i$ is the extension of $\hat A^i$ to the bulk. In order to make contact with the 5d theory, one takes the seven-dimensional space to contain a $S^1$ factor in the following way $\c M^7 = \c N_6 \times S^1$, with $\partial \c N_6 = \c M_5$, and expends $A^i_{(7)} = A^i_{(6)} + \zeta^i_{(6)} (\d y -A^0_{(6)})$, where
the subscript ${(6)} $ denotes the extensions of the 5d fields to the bulk $ \c N_6 $.
For constant $\zeta^i_{(6)} = \big <{\zeta^i_{(6)}} \big >$, the action \eqref{CS7} becomes a total derivative on $\c N_6$, such that one finds the following contribution in 5d
\begin{equation}\label{AI}
S_5^{\rm inflow} = \,\mfrac{2}{3} \, \sum_\f  \, q^\f_i \,q^\f_j\, q^\f_k\, q^\f_l \int_{\c M_5} \vev{\zeta^i} \, \widetilde A^j \, \widetilde F^k \,  \widetilde F^l \,,
\end{equation}
with $\wt A^i \equiv \frac 1c \, ( A^i - \vev{\zeta^i} A^0)$. This term is precisely the difference between the two regularizations we used, $S_5^{\rm inflow} = S_5^{\reg} -  S_5^{\regb} \equiv \Delta S^{\reg}$ . 
In order to see why this can happen, the key is to note that the zeta function regularization is by construction invariant under the shifts $\vev{\zeta^i} \to \vev{\zeta^i} +1$. These shifts originate from 6d large gauge transformations along the circle an since the other regularization correctly reproduces the anomaly, all the non invariance in 5d is encoded in $ \Delta S^{\reg} $. At the same time we saw that the all the non invariance of the 6d 1PI action is captured by the 7d CS term~\eqref{CS7}, such that it is natural that its reduction gives $ \Delta S^{\reg} $. Actually the same argument holds for non-constant $\zeta^i$'s, considering also the small gauge transformations $\delta A^i = \d \lambda^i$. Although in that case the action~\eqref{CS7} does not give a total derivative on $\c N_6$, its variation still does, and this correctly reproduces the variation of $ \Delta S^{\reg} $ (i.e.~of $- S^\regb_5 $ since $ S^\reg_5 $ is invariant).

Let us now assume that we started from a consistent 6d (1,0) theory, i.e with eqs.~\eqref{AC_6d} holding. We already saw that the non-invariant terms~\eqref{ngi_all} cancel because of~\eqref{gauge}. Additionally, using the other anomaly equations~\eqref{grav_irr} to~\eqref{mixed}, we find the following total CS terms (summing the classical and the one-loop pieces)
\begin{equation}\label{CS_gi_full}
\begin{split}
\c L^{\rm \textsc{cs},\, quant} 
 = - \mmfrac 14 A^0  F  \! \inner  F + \mmfrac 14 \, b_{ij} \inner A   F^i F^j 
 \msplit{65}
 - \mmfrac{1}{24} \sum_{\f}   q^\f_i \,q^\f_j\, q^\f_k \,   A^i F^j F^k  - \mmfrac{1}{8}  \, a \inner b_{ij}  A^i F^j F^0 -\mmfrac{1}{48}  \,  a \inner a  \, A^0 F^0 F^0\,.
\end{split}
\end{equation}
In this expression, we also assumed that $\vev{\zeta^i} \ll 1$ for all $i$, such that $\floor{\mu^\f} = 0$ and $S_k (\mu^\f) =0$ for all $k$, since this is the most straightforward case to match with the M-theory reduction on a Calabi-Yau threefold, as we will see in next section. However relaxing this assumption is straightforward: it shifts in Chern-Simons levels by integers. 
In section~\ref{sec:Mthy}, we will give a geometric interpretation of this fact in F-theory.  
\note{Lift to 6d cf Bonetti/Kapfer.}

The CS terms~\eqref{CS_gi_full} are gauge invariant and can thus be obtained in the canonical manner
from the prepotential
\begin{equation}\label{NF}
\begin{split}
\cubN^{\rm quant}_{S^1} = \mmfrac 12 \, \phi^0  \phi \inner  \phi - \mmfrac 12  \, b_{ij} \inner   \phi  \,  \phi^i \phi^j 
+ \mfrac  1{12} \sum_{\f}   q^\f_i \,q^\f_j\, q^\f_k \,  \phi^i  \phi^j  \phi^k + \mfrac  1{4} \, a \inner b_{ij} \,  \phi^0  \phi^i  \phi^j + \mfrac  1{24} \, a \inner a \, ( \phi^0)^3 \,.
\end{split}
\end{equation}
Not only will this formulation will be more convenient for the match with the M-theory reduction in the next section, but it also allows us to infer the quantum corrections to the kinetic terms~\eqref{Lkin}.
Indeed this corrected potential is perturbatively exact, as CS terms receive only one-loop corrections, such that one can compute the corrected kinetic terms using the general formula~\eqref{gcf}. 

However one should first remember that the coordinates~\eqref{very_spec_coord} were constructed in such a way that the classical prepotential $\cubN^{\rm class}_{S^1}$ in~\eqref{Nclass} equates one.
Taking quantum corrections into account (including those for the kinetic terms) does not break supersymmetry, and the quantum corrected prepotential~\eqref{NF} should still equate to one. This will also be clear later on when we match with the M-Theory set-up, as it will match with $\cubN_M=1$, defined in eq.~\eqref{prepot}. 
For $\cubN^{\rm quant}_{S^1}=1$ to hold, the coordinates~\eqref{very_spec_coord} should also receive quantum corrections. A simple possibility is the following
\begin{subequations} \label{very_spec_coord_new}
	\begin{alignat}{3}	
	    &\phi ^0 &&= \,\const  \,            && r^{-4/3}  \,, \\
		&\phi ^\a  &&=  \frac{\const}{4}\, && r^{2/3} \,\Big (\sqrt{1 -  \Delta} \, j^\a + 2 \, r^{-2} \, b^\a_{ij} \, \zeta^i \zeta^j \Big ) \,, \\
		&\phi ^i &&= \, \const \,             && r^{-4/3} \, \zeta^i \,,
	\end{alignat}
\end{subequations}
where all the information about the corrections is encoded in the function
\begin{align}\label{key}
\Delta  &= 32 \, r^{-4} \Big(\mmfrac 1{24} \, a \inner a + \mfrac 14 a \inner b_{ij} \,\zeta^i \zeta^j + \mmfrac 1{12} \sum_\f q^\f_i \,q^\f_j\, q^\f_k \, \zeta^i \zeta^j \zeta^k - \mmfrac{1}{24} \, b_{ijkl} \,\zeta^i \zeta^j \zeta^k \zeta^l\Big) \,.
\end{align}
Since the scalars are real, we must have $\Delta < 1$; this can certainly be achieved by taking $r$ sufficiently large. 

With this identification of the coordinates and the prepotential~\eqref{NF}, we can compute the quantum corrections to the kinetic terms using~\eqref{gcf}. The result is 
\begin{equation}\label{Lkinquant}
\begin{split}
\c L ^{\rm kin,\, quant} = \ & -\mfrac 23 \, r^{-2} \, \frac{1-4  \Delta}{1-\Delta} \, \d r \ws \d r + r^{8/3} \Big (-\mfrac 14 + \mfrac 12 \, \Delta - \Delta^2 \Big ) \, {\wt F^0}\ws \wt F^0 \\
& - \Big ( \mfrac 12 - \mfrac 32 \, \Delta + \Delta^2 \Big) \, g_{\a\b} \, \d j^\a \ws \d j^\b 
+ r^{-4/3} \Big (- \mfrac 12 + \Delta \Big )  \, g_{\a\b} \, {\wt F}^\alpha \ws \wt F^\b
 \\[1mm] 
&-\Bigg[ 2\, r^{-2} \, j \inner b_{ij}  \, \sqrt{1-\Delta} - \mfrac{1}{16} \frac{   (1-4  \Delta) \, \Delta_{ij} +2 \, \Delta_i  \Delta_j }{ 1-\Delta} \Bigg] \d \zeta^i \ws \d \zeta^j \\
&- \bigg[  2\, r^{2/3} \, j \inner b_{ij} \, \sqrt{1-\Delta}  - \mfrac 14 \, r^{8/3} \, \Big(\Delta_{ij} - \Delta_i  \Delta_j \Big)\bigg] \, \wt F^i \ws \wt F^j \\
&-r^{-1} \frac{ \Delta_i}{1-\Delta} \d r \ws \d \zeta^i+  r^{8/3} \,\Big ( \mmfrac 12 - \Delta  \Big ) \, \Delta_i \, \wt F^0 \ws \wt F^i \\
&+\, r^{2/3} \,j_\a\, \sqrt{1-\Delta}  \, \bigg( \Delta_i  \,  \wt F^i \ws \wt F^\a -2 \, \Delta \, \wt F^0 \ws \wt F^\a  \bigg)\,,
\end{split}
\end{equation}
where $\Delta_i \equiv \partial_{\zeta^i}\Delta$ and the field strengths $\wt F$ are defined as in~\eqref{curlyF}.
The expression~\eqref{Lkinquant} takes into account all the corrections to the kinetic terms for the circle reduction of the action~\eqref{6Dact}.

\section{M-theory on a Calabi-Yau threefold}
\label{sec:Mthy}

In this section, we present the 5d effective action of M-theory on a Calabi-Yau threefold, obtained from the reduction of the eleven-dimensional supergravity~\cite{Cadavid:1995bk}.\footnote{Higher derivative terms from M-theory have been computed~\cite{Antoniadis:1997eg} and can be lifted in F-theory, see e.g.~\cite{Bonetti:2011mw,Grimm:2013oga}, but since they are not relevant for our analysis, we will work at the two derivatives level.}
For this reduction to be dual to an F-theory one, the Calabi-Yau needs to be elliptically fibered. In addition, we will restrict ourselves to smooth Calabi-Yau's with extra sections, as this corresponds to the 6d theories considered in the previous section: abelian gauge theories arise from singularities while abelian ones from extra sections~\cite{Morrison:1996na,Morrison:1996pp}.
We denote the projection onto the base by $\pi: Y_3 \to B_2$ and the sections by $\s_i, \ i=0,...,r-i$ where $r$ is the rank of the Mordell-Weil group of the Calabi-Yau. Finally, we will focus only on the vector sector, omitting hypermultiplet sector.
\note{What if no zero section?}

\subsection{Classical reduction}

The 5d vectors arise from expanding the three-form $\hat C_3$ (the hat now indicates that it is eleven-dimensional) along harmonic (1,1)-forms $\omega_I, \ I=1,...,\hoo{Y_3}$,
\begin{equation}\label{key}
\hat C_3 = A^I \wedge \omega_I + ... \,.
\end{equation}
One thus finds $h^{1,1}(Y_3)$ vectors $A^I$, of which one resides in the 5d gravity multiplet, such that there are $\nvf  = h^{1,1} (Y_3)-1$ vector multiplets in 5d. 

Concerning the scalars of the vector multiplets, they arise as Kähler structure deformations, which are obtained by expanding the K\"ahler form $J$ along the two-forms~$\o_I$ as~$J=  v^I  \omega_I$. This seems to lead to one scalar too many. However, the overall volume of the Calabi-Yau threefold
\begin{equation}\label{volumeY3}
\c V = \mmfrac{1}{3!} \int_{Y_3 } J\wedge J \wedge J = \mmfrac{1}{3!} \, \c K_{IJK} \, v^I v^J v^K \,,
\end{equation}
actually sits in a hypermultiplet. 
In this expression, the coefficients $\K{IJK}$ are the intersection numbers of the Calabi-Yau, given by
\begin{equation}\label{intnbY3}
\c K _{IJK}   = \int_{Y_3} \o_I \wedge \o_J \wedge \o_K= D_I \cap D_J \cap D_K\,.
\end{equation}
The $D_I$'s are the divisors Poincaré dual to the two-forms~$\o_I$. 
In order to separate the total volume $\cV$ from  the scalars of the vector multiplets, it is natural to define the latter as
\begin{equation}\label{LI-def}
\varphi^I = \frac{ v^I}{\c V^{1/3}}\ .
\end{equation}
These scalars parametrize only $h^{1,1}(Y_3)-1$ degrees of freedom, as it should, since they satisfy
\begin{equation}\label{prepot}
\cubN_M \equiv \mfrac{1}{3!} \, { \c K}_{IJK} \varphi^I \varphi^J \varphi^K =1\ .
\end{equation}
This condition matches with the very-special K\"ahler constraint~\eqref{constr}, suggesting that the $\varphi^I$'s are to be identified with the 5d the very-special coordinates, and that $\cubN_M$ is the prepotential.

This is indeed the case, as the 5d CS terms, which arise from the 11d CS term, are given by 
\begin{equation}\label{key}
S^\cs = -\mfrac 1{12} \int_{\c M_5 \times Y_3} C_3 \wedge G_4 \wedge G_4 = - \mfrac{1}{12} \, \K{IJK} \, \int_{\c M_{5}} A^I \wedge F^J \wedge F^k\,,
\end{equation}
while the metric is found~\cite{Cadavid:1995bk} to be \begin{equation}\label{key}
G_{IJ} = \mfrac 12 \int_{Y_3} \o_I \ws \, \o_J = \left . -\mfrac 12\, \partial_I \partial_J (\log \c F_M) \right |_{\c F_M =1}\,.
\end{equation}

\subsection{F-theory frame}

Up to here the discussion was for a generic Calabi-Yau threefold, we now generalize to the case of smooth elliptic fibration with extra sections. In such cases there are several types of (1,1)-form, and corresponding divisors. 
The appropriate basis of divisors $D_I=(D_0,D_\a,D_i)$ for lifting to F-theory consists of the following~\cite{Bonetti:2011mw,Grimm:2013oga}:
\begin{itemize}
\item The divisor $D_0$ obtained from the zero section as 
\begin{equation}\label{shiftD0}
D_0 = \s_0 - \mmfrac 12 \, (\s_0 \cap \s_0 \cap D^\a) \, D_\a \,,
\end{equation}
which ensures that $D_0 \cap D_0 \cap D_\a =0$.
The associated vector corresponds to the KK-photon $A^0$.
\item The vertical divisors $D_\a$, which are pullbacks of the base divisors~$D_\a^\rb$, 
\begin{equation}\label{key}
D_\a = \pi^* (D^\rb_\a), \qquad \a = 1, ..., \hoo{B_2}\,.
\end{equation}
The associated vectors $A^\a $ lift to the tensors $\hat B^\a$ in 6d. 
\item The divisors $ D_i, \ i=1,...,r $ obtained from the extra sections $\sigma_i$ through Shioda map 
\begin{equation}\label{Shioda}
D_i = \s_i - \s_0 - \Big [(\s_i- \s_0) \cap \s_i \cap D^\a \Big ] \, D_\a\,.
\end{equation}
The associated vectors $A^i$ lift to the $ U(1) $ vectors $\hat A^i$ in 6d.
\end{itemize}
In this basis, one can work out some of the intersection numbers in~\eqref{intnbY3}
\begin{equation}\label{intnumb_rat}
\begin{alignedat}{4}
\K{\a\b\g}  & = 0 \,,  & \K{\a\b i} & = 0 \,,  \\
\K{00\a}  & = 0 \,,  & \K{0 \a i} & = 0\,,\\ 
\K{0\a\b} & = \eta_{\a\b} \,, \qquad & \K{\a ij} & = \pi(D_i \cap D_j)_\a   \,, 
\end{alignedat}
\end{equation}
where $\eta_{\a\b} = D^\rb_\a \cap D^\rb_\b $ is the intersection matrix on the base and $\pi(\S) = (\S \cap D^\a) D_\a^\rb$ is the projection of a two cycle $\S$ on the base. The metric $\eta_{\a\b}$ can be used to raise and lower the $\a$-indices, and we will denote by $\innerY$ the product made using this metric, $v \innerY w \equiv \eta_{\a\b} v^\a w^\b$.

When the zero section $\s_0$ is holomorphic, one in addition finds 
\begin{equation}\label{intnumb_hol}
\begin{aligned}
\K{000} & = \mmfrac 14 \, K \innerY K  \,, \\[1mm]
\K{00i} & = 0  \,,  \\
\K{0ij} & = -\mmfrac 12 \, \pi(D_i \cap D_j) \innerY K \,,
\end{aligned}
\end{equation}
where $K^\a$ are the coefficients in the expansion of the anti-canonical class of the base, i.e.~$K^{-1}_{B_2} = K^\a D^\rb_\a$.

Expanding the Kähler form in the (1,1)-form basis Poincaré dual to $(D_0,D_\a,D_i)$, and rescaling as in~\eqref{LI-def}, one obtains the split  $\varphi^I=(\varphi^0,\varphi^\a,\varphi^i)$ of the vector multiplets scalars. Using the intersection numbers~\eqref{intnumb_rat} and~\eqref{intnumb_hol}, the prepotential~\eqref{prepot} then reads 
\begin{equation}\label{NM}
\begin{split}
\cubN_M =  \ \mmfrac 12 \, \varphi^0 \,\varphi \innerY \varphi + \mmfrac 12 \, \pi(D_i\cap D_j) \innerY \varphi \, \varphi^i \varphi^j \msplit{56} 
+ \mmfrac 16 \, \K{ijk} \, \varphi^i \varphi^j \varphi^k 
-\mmfrac 14 \, \pi(D_i\cap D_i) \innerY K \,  \varphi^0 \, \varphi^i \varphi^j   + \mmfrac{1}{24} \, K \innerY K \, \big (\varphi^0\big )^3\,.
\end{split}
\end{equation}
As we will see in next subsection, this basis is the correct one for lifting to F-theory.

\subsection{Match with the circle reduction}

As explained earlier, the M/F-duality tells us that reduction of M-theory on an elliptically fibered smooth CY$_3$
is to be matched with the circle reduction of a 6d (1,0) theory, pushed on the Coulomb branch and with the massive mode integrated out (quantum corrections being taken into account).
We stress that the two theories being matched are very different in nature: one is a classical reduction, meaning that only the zero-modes are kept, while the other one is a quantum corrected action, with the massive modes running in the loops.

In particular it implies that the prepotentials~\eqref{NM} and~\eqref{NF} should match. This is indeed the case if one makes the following identifications. First, one identifies the coordinates and metric,
\begin{align}\label{key}
\phi^I&=\varphi^I \,, \\
\eta_{\a\b} & = \Omega_{\a\b} \,.
\end{align}
Second, the anomaly coefficients are identified as
\begin{equation} \label{an_coeff}
\begin{aligned}
a^\a & = K^\a\,,\\
b^\a_{ij} & = - \pi (D_i \cap D_j)^\a \,,
\end{aligned}
\end{equation}
which agrees with the findings of~\cite{Sadov:1996zm} and~\cite{Park:2011ji} respectively.\footnote{In fact imposing that these identifications hold, the match of two prepotentials leaves no room for relative coefficients in the coordinates, hence the \emph{à priori} uncanny factors in the definition~\eqref{very_spec_coord} of the coordinates.} 
Finally the triple intersection numbers $\K{ijk}$ are to be matched with the sum of the cube of the charges,
\begin{equation}\label{Kijk}
{\c K}_{ijk} = \mmfrac 12 \sum_\f q_i^\f q_j^\f q_k^\f\,,
\end{equation}
which was already noted and checked for several examples in~\cite{Grimm:2013oga}.
It is worth noting that for the match to be possible, the pure gravitational and mixed gauge-gravitational anomalies also need to be canceled, as was also noted in~\cite{Grimm:2013oga}.

When the zero section is not holomorphic, eqs.~\eqref{intnumb_hol} do not hold anymore, such that the prepotential differs from~\eqref{NM} and the match with the circle reduction seems compromised. It was noted in~\cite{Grimm:2013oga} that in such a case, the assumption $\vev{\zeta^i} \ll 1$ cannot be made for all $i$, meaning that the functions $S_n(\mu^f)$ in~\eqref{PV} do not vanish, and the match with the M-theory reduction is still possible.

\note{Match only possible if the mixed and gravitational anomalies also cancel. Refer to Federico and Thomas.}

\section{Conclusion}

In the first part of this note, we showed that the pure abelian anomaly of 6d (1,0) theories is preserved upon circle reduction, showing up as field-dependent Chern-Simons terms in the 5d effective action. These terms vanish if and only if the 6d anomaly is canceled, which requires a Green-Schwarz mechanism. They appear at one-loop upon integrating out the massive KK-towers of hyperini. This yields at first a divergent result which needs to be regularized, and we showed that the result depends on the regularization. In order to correctly reproduce the anomaly (and in particular have no such field-depend CS terms when it is canceled) we argued that it is crucial to use a regularization that preserves 6d Lorentz invariance, as these are the symmetries of the UV of the theory. 
In this way we found the correct cubic prepotential of the 5d \susy{1} theory, which is exact as the CS terms do not receive corrections beyond one-loop. From this prepotential we inferred the quantum corrections to the kinetic terms.

In the second part of this note, we matched this prepotential with the one obtained by reducing M-theory on a Calabi-Yau threefold. A perfect match can be achieved if and only if the 6d theory was anomaly free; in particular, the M-theory effective action does not contain field-dependent Chern-Simons terms. This means that 6d (1,0) F-theory compactifications obtained via M-theory are necessarily anomaly free, as is expected.

For simplicity we restricted our analysis to 6d supergravities with abelian gauge groups, as the in 5d Coulomb branch non-abelian gauge groups are anyway broken down to their Cartan $ U(1) $. It would nevertheless be interesting to extend our analysis to non-abelian gauge groups, as well as to other kinds of anomalies, like mixed or pure gravitational anomalies, which are likely to be related to mixed or pure gravitational CS terms in 5d, or to anomalies of the global part of the gauge group, which cannot be accessed via Feynman diagrams and thus would require a different treatment than the one presented here. 

It would also be interesting to relate our understanding of 6d anomalies to the recent work~\cite{Katz:2020ewz} where some 5d $\c N=1$ theories are shown to belong to the swampland by studying the anomaly inflow on the worldsheet of the strings which source the two-form fields. If those strings also induce field-dependent CS terms in the bulk, our analysis would imply that they cannot originate from consistent 6d theories.

\section*{Acknowledgments}

It is a pleasure to thank Federico Bonetti, Markus Dierigl, Iñaki Garc\'ia Etxebarria, Thomas Grimm, Miguel Montero and Irene Valenzuela for valuable discussions and comments on the draft.
This work is supported by the Spanish Research Agency (Agencia Estatal de Investigaci\'on) through the grant IFT Centro de Excelencia Severo Ochoa SEV-2016-0597, and by the grant PGC2018-095976-B-C21 from MCIU/AEI/FEDER, UE.

\appendix

\section{Chern-Simons terms from integrating out KK-towers}
\label{app:loop}

In this appendix, we compute the CS terms generated by the diagram in figure~\ref{fig:aaa5d-no-text} in the main text, 
for a KK-tower of a 6d spin-1/2 fermion running in the loop. 
As mentioned in the main text, the amplitude diverges, and can be regularized in different ways. The most straightforward is the zeta function regularization, which give the result~\eqref{zetaijk}. Here we will use the Pauli-Villars (PV) regularization, which consists in adding a fermion of opposite chirality with mass $M$ and at the end send $M\to \infty$. This can be achieved in two very different ways, presented in subsections~\ref{sec:reg} and~\ref{sec:regb} respectively. The first one consists in adding one PV regulator per KK-mode, making no reference to the six-dimensional origin of the theory and therefore preserving only 5d Lorentz invariance. The second comes from the reduction of a PV regulator of the 6d theory and consequently intrinsically preserves 6d Lorentz invariance, while explicitly breaking gauge invariance. As explained in the main text, the second approach is the appropriate one if the theory originates from a circle reduction, for regularization is a process taking place in the UV of the theory.

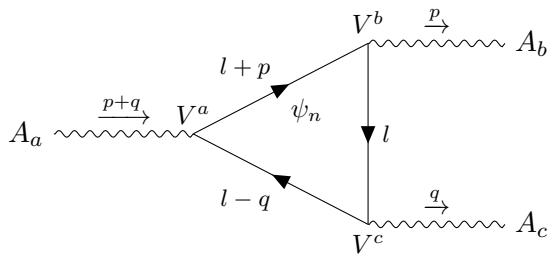
\begin{figure}[h!]
\centering
\begin{tikzpicture}
\begin{feynman}
\vertex (aa) {\(A_a\)} ; 
\vertex [right=2.2 cm of aa, label ={\small \( V^a\)}] (a); 
\vertex [right=2.3cm of a] (ai); 
\vertex [above= 1.2cm of ai, label ={\small \( V^b\)}] (b);
\vertex [below=1.2 cm of ai, label =-90:{\small \( V^c\)}] (c);
\vertex [right=1.8 cm of b] (ab) {\(A_b\)};
\vertex [right=1.8 cm of c] (ac) {\(A_c\)};
\diagram* {
(aa) -- [photon, edge label ={\small \( \xrightarrow{p+q}\)}] (a),
(a) -- [fermion, edge label= {\footnotesize \(l+p\)}, edge label' ={\small \( \psi _n\)}] (b) -- [fermion,edge label= {\footnotesize \(\, l\)}] (c) -- [fermion, edge label= {\footnotesize \(l-q\)}] (a), 
(b) -- [photon, edge label = {\small \( \xrightarrow{p}\)}] (ab),
(c) -- [photon, edge label = {\small \( \xrightarrow{q}\)}] (ac),
};
\end{feynman}
\end{tikzpicture}
  \captionsetup{width=.49\linewidth}
	\caption{The AAA diagram in 5d, with external momenta and polarizations specified.}
	\label{fig:aaa5d}
\end{figure}

For simplicity, we will restrict our discussion to the case where all external gauge fields belong to vector multiplets, i.e.~on the coefficients $k_{ijk}$. Additionally, we carry out the computation for a single KK-tower running in the loop and for a unique external gauge field, but the generalization being is straightforward. 
With the external momenta and polarizations fixed as in figure~\ref{fig:aaa5d}, the amplitude is given by 
\begin{equation}\label{diag}
\Delta^{abc} = \sum_{n \in \Z} \intdkn 5 \, \tr \Big[{V^a \, \c S_n({l+p}) \, V^b \, \c S_n({l}) \,  V^c \, \c S_n({l-q})}\Big],
\end{equation}
where we denoted by  $\c S_n(l)$ the propagator of the $n$-th fermionic KK-mode with momentum $l$ and by $V^a$ the interaction vertices.
This leads a contribution to the Wilsonian effective action proportional to $\Delta^{abc} A_a A_b A_c$. We are interested in the parity violating part, i.e.~the $\e$ part of the amplitude. More precisely, we are interested in the CS terms $\frac{1}{12} \, k \, AFF$, which are induced by $\Delta^{abc} \propto i\,k\, \epsilon^{abcde} p_d q_e$.

In order to make the comparison between the two regularizations easier, we will use for both a ``six-dimensional formalism'', meaning that we take the $\g$-matrices to be $8 \times 8 $. In this formalism, the propagator and vertex of the $n$-th KK-mode of a charged hyperino read 
\begin{equation}\label{key}
\c S_n(l) = P_L \, \frac{1}{\gamma^a  l_a - \gamma^5 m_n } , \qquad V^a = q \gamma^a P_L\,.
\end{equation}
where $P_L$ is the chirality projector $P_L = \frac 12 (1+ \g^*)$, with $\g^* = \g^0 ... \g^5$, and
\begin{equation}\label{key}
m_n = \mkk + \mcb = \mfrac{1}{r} \left(\mfrac{r_0}{r}\right)^{1/3} \Big (n + q \, \vev{\zeta}\Big)
\end{equation}
is the total mass as in~\eqref{mn}.

\subsection{Regularization preserving 5d Lorentz invariance}
\label{sec:reg}

\note{Add the actions and their circle reductions?}

As explained above, we first regularize the sum in~\eqref{diag} by adding one 5d PV regulator per KK-mode. The propagators of these regulators read 
\begin{equation}\label{key}
\c S_n^\reg(l) = P_L \,\frac{1}{\g^a l_a - \g^{5} m_n -M} \,.
\end{equation}
Plugging back in~\eqref{diag} one finds
\begin{align}
k^\reg & = q^3 \limm \sumn\intl  \bigg(\frac{m_n}{(l^2 - m_n^2)^3} - \frac{m_n}{(l^2 - m_n^2-M^2)^3}\bigg)\,, \notag  \\
& =  q^3  \limm \frac 12 \sum_{n\in \Z } \Bigg ( \sign(m_n) - \frac{m_n}{\sqrt{m_n^2+M^2}} \Bigg) \,, \notag \\
& = q^3 \, \Big( \mfrac 12 + \floor{q \, \vev{\zeta}} - q \, \vev{\zeta} \Big) \,,
\end{align}
where in the second line we performed a Wick rotation. This purely 5d regularization thus yields the same result as the zeta-function one~\eqref{zetaijk}.

\subsection{Regularization preserving 6d Lorentz invariance}
\label{sec:regb}

Alternatively, one can add a single PV regulator in 6d and reduce it on the circle. Since the 6d fermions are chiral, adding a massive PV regulator necessarily breaks gauge invariance, and thus also do its KK-modes. Their propagators read (see~\cite{Corvilain:2017luj} for more details)
\begin{equation}\label{Sregb}
\c S_n^\regb(l) = \frac{1}{\g^a l_a - \g^{5} \big(\mkk + \mcb P_L \big) -M} \,.
\end{equation}
In order to compute the trace in~\eqref{diag} we need to invert the matrix in~\eqref{Sregb}, and since this regulator is non-standard, let us give the result here 
\begin{equation}\label{key}
\c S^\regb_n(l) =  \frac 1{D_l}\left(\g^a l_a - \g^5 (\mkk + \mcb P_R) + M  \right)\left( a_l - \mcb\Big( ( \mcb + 2 \mkk  ) P_L  M \g^5 \g^*\Big) \right) \,
\end{equation}
where we defined
\begin{equation}
	\begin{aligned}
    D_l  &= a_l b_l - \mcb^2 M^2 \,,\\
	a_l  &= l^2 - \mkk^2 - M^2\,,\\
	b_l  &= l^2 - \mn^2 -M^2\,.
	\end{aligned}
\end{equation}
One then finds 
\begin{align}\label{key}
k^\regb &= q^3 \lim_{M\to \infty} \sum_{n \in \Z}\intl \Bigg ( \frac{m_n}{(l^2- m_n^2)^3} - \frac{a_l}{D_{l}^3} \bigg [ 
a_{l}^2 m_n + \mcb M^2 \Big (  a_l + \mfrac 45 \, l^2 \Big ) \bigg ] \Bigg ) \,, \notag \\
& =  q^3 \, \Big(\mmfrac 12 + \floor{q \, \vev{\zeta} } - \mfrac 34 \, q \, \vev{\zeta} \Big ) \,.
\end{align}
This is the result we presented in~\eqref{PVijk}, when including several KK-towers running in the loop and different external gauge fields.

\bibliographystyle{JHEP}
\bibliography{anomalies.bib}

\end{document}